\documentclass{aa}
\usepackage{graphicx}
\usepackage{bm,natbib,url,wasysym}
\usepackage[varg]{txfonts}
\usepackage{xcolor}
\usepackage{amsmath}
\bibpunct{(}{)}{;}{a}{}{,} % to follow the A&A style
\graphicspath{{./figs/}{./png/}}
\newcommand{\brac}[1]{\langle #1 \rangle}
\newcommand{\EQ}{\begin{equation}}
\newcommand{\EN}{\end{equation}}
\newcommand{\EQA}{\begin{eqnarray}}
\newcommand{\ENA}{\end{eqnarray}}

\newcommand{\Eq}[1]{Equation~(\ref{#1})}
\newcommand{\Eqs}[2]{Equations~(\ref{#1}) and~(\ref{#2})}

\newcommand{\Sec}[1]{Section~\ref{#1}}

\newcommand{\Fig}[1]{Fig.~\ref{#1}}

\newcommand{\Tab}[1]{Table~\ref{#1}}

\newcommand{\bra}[1]{\langle #1\rangle}

\newcommand{\fluc}[1]{#1^\prime}

{}
{}
{}

\newcommand{\meanuu}{\overline{\mbox{\boldmath $u$}}{}}{}
{}
{}
{}
{}
{}
{}
{}
{}
\newcommand{\meanBB}{\overline{\mbox{\boldmath $B$}}{}}{}
{}
{}
{}
{}
{}
{}
{}
{}

{}
{}
{}

\newcommand{\meanB}{\overline{B}}

{}
{}
{}
{}

{}

{}
{}

\newcommand{\alphapp}{\alpha_{\phi\phi}}

\newcommand{\Ot}{\tilde{\Omega}}

\newcommand{\uu}{\mbox{\boldmath $u$} {}}

\newcommand{\bb}{\mbox{\boldmath $b$} {}}
\newcommand{\BB}{\mbox{\boldmath $B$} {}}

\newcommand{\AAA}{\mbox{\boldmath $A$} {}}

\newcommand{\nab}{\mbox{\boldmath $\nabla$} {}}
\def\Ta{\mbox{\rm Ta}}
\def\Ra{\mbox{\rm Ra}}

\def\Co{\mbox{\rm Co}}
\def\Ro{\mbox{\rm Ro}}

\def\Roo{\mbox{\rm Ro}^{-1}}

\def\PrSGS{\mbox{\rm Pr}_{\rm SGS}}

\def\Pm{\mbox{\rm Pr}_{\rm M}}
\def\Rm{\mbox{\rm Re}_{\rm M}}

\def\Rey{\mbox{\rm Re}}

\def\Pe{\mbox{\rm Pe}}
\def\Co{\mbox{\rm Co}}

\def\kf{k_{\rm f}}

\def\urms{u_{\rm rms}}

\def\etat{\eta_{\rm t}}

\newcommand{\s}{\,{\rm s}}

\newcommand{\m}{\,{\rm m}}

\newcommand{\mpsec}{\,{\rm m/s}}

\newcommand{\kg}{\,{\rm kg}}

\newcommand{\Mm}{\,{\rm Mm}}

\newcommand{\chiS}{\chi_{\rm SGS}}
\newcommand{\chiSm}{\chi_{\rm m}^{\rm SGS}}

\begin{document}

\titlerunning{Dynamo cycles in global convection simulation}
\authorrunning{Warnecke}

\title{Dynamo cycles in global convection simulations of
  solar-like stars}
\author{J. Warnecke\inst{1}}
\institute{Max-Planck-Institut für Sonnensystemforschung,
  Justus-von-Liebig-Weg 3, D-37077 G\"ottingen, Germany\\
\email{warnecke@mps.mpg.de}\label{inst1} 
}

%\date{\today,~ $ $Revision: 1.12 $ $}
\abstract{
Several solar-like stars exhibit cyclic magnetic activity similar to
the Sun as found in photospheric and chromospheric emission.
}{%
We want to understand the physical mechanism involved in rotational dependence of these activity
cycle periods.
}{%
We use three-dimensional magnetohydrodynamical simulations of global
convective dynamos models of solar-like stars to investigate the rotational dependency of
dynamos. We further apply the test-field method to determine the $\alpha$
effect in these simulations.
}{%
We find dynamo with clear oscillating mean magnetic fields for
moderately and rapidly rotating runs. For slower rotation, the field is
constant or exhibit irregular cycles. In the moderately and rapidly
rotating regime the cycle periods increase weakly with rotation. This
behavior can be well explained with a Parker-Yoshimura dynamo wave
traveling equatorward. Even though the $\alpha$ effect becomes
stronger for increasing rotation, the shear decreases steeper, causing
this weak dependence on rotation. Similar as other numerical studies,
we find no indication of activity branches as suggested by
\cite{BST1998}. However, our simulation seems to agree more with the
transitional branch suggested by \cite{DLLMS17} and \cite{OLKPG17}.
If the Sun exhibit a dynamo wave similar as we find in our
simulations, it would operate deep inside the convection zone.
}{}

\maketitle

\section{Introduction}
The Sun, our nearest late-type star exhibits a magnetic activity
cycle with a period of around 11 yrs.
The cyclic magnetic field is generated by a dynamo operating below the
surface, where it converts the energy of rotating convective turbulence
into magnetic energy.
The solar dynamo mechanism is still far from being fully understood
\citep[e.g.][]{O03,Ch14}.
One reason is the limited information about the dynamics in the solar
convection zone.
Helioseismology have provided us with the profile of temperature and
density stratification and the differential rotation
\citep[e.g.][]{Schouea98} in the interior, further information such as
the meridional circulation profiles, convective velocity strength or even magnetic
field distributions are currently inconclusive or not even possible \citep[e.g.][]{Basu:2016,HGS16}.
One way to investigate how important differential rotation, meridional
circulation and turbulent convective velocities are for the solar
dynamo, is to use numerical simulations.
Since the early simulations by \cite{G83}, there have been several
advances using numerical simulation due to the increase of computing resources.
Nowadays, global simulations of convective dynamos are able to reproduce
cyclic magnetic fields and dynamo solutions resembling many features of
the solar magnetic field evolution \citep{GCS10,KMB12,WKKB14,ABMT15},
even the long-time evolution \citep{ABMT15,KKOBWKP16,BSCC16}.
The cyclic magnetic field in these simulations can be well understood in
terms of Parker-Yoshimura rule \citep{P55,Yos75, WKKB14}, where a propagating
$\alpha\Omega$ dynamo wave is excited, see also \cite{GDW12}.
The $\alpha$ effect \citep{SKR66} describes the magnetic field enhancement from helical
turbulence and the $\Omega$ effect the shearing of magnetic field
caused by the differential rotation.
The propagation direction of the dynamo wave depends on the sign of
$\alpha$ and shear: for generating an equatorward propagating wave, the
product of $\alpha$ and the radial gradient of $\Omega$ must be
negative (positive) in the northern (southern) hemisphere. 
For explaining the solar equatorward propagation of the sunspot
appearance by the Parker-Yoshimura rule therefore requires either to invoke the
near-surface-shear layer \citep{B05}, because only there the radial
gradient is negative \citep{BSG14} and $\alpha$ is positive or that
$\alpha$ changes sign in the bulk of the convection zone
\citep{DWBF15}, where the radial shear is positive.
Furthermore, to fully understand the magnetic field evolution in the
global numerical simulation one needs suitable analysis tools to
extract the important contribution of turbulent dynamo effects.
One of these tools is the test-field method
\citep{SRSRC05,SRSRC07,WRTKKB17}.
This method allows to determine the turbulent transport coefficients
directly from the simulations. This include the measurement of
tensorial coefficients such as $\alpha$, turbulent pumping and
turbulent diffusion.
Already the first application to global convection simulation of
solar-like dynamo revealed that the turbulent effects can have a
significant impact on the large-scale magnetic field dynamics
\citep{WRTKKB17,GKW17}.

Another possibility to understand the solar dynamo make use of the
observation of other stars.
Since the Mount Wilson survey, we know that many stars exhibit
cyclic magnetic activity
\citep[e.g.][]{Noyes+al:1984b,Noyes+al:1984a,Baliunas1995}. In this
survey, they observe solar-like stars
in the chromospheric Ca II H\&K band, which is used as a proxy for
magnetic activity.
Using this data \cite{BST1998} and \cite{SB1999} found two distinguish branches,
when they plot the ratio of rotational period and activity cycle
period $P_{\rm rot}/P_{\rm cycl}$ over the rotational influence on the stellar convection in terms of
the inverse Rossby number. The two branches are called inactive and
active branch, because of their preferred magnetic activity, divided by
the so-called Vaughan--Preston gap \citep{VP1980}. Their slopes are positive in
terms of rotational influence, this means the cycle period decreases
faster than linear with increasing rotation.
This agrees qualitatively with the finding of \cite{Noyes+al:1984a}, where
they obtain $P_{\rm cycl} \propto P_{\rm rot}^{1.25}$.
However, \cite{OKPSBKV16} find in their recent re-analysis of the Mount
Wilson data a relation of $P_{\rm cycl} \propto P_{\rm rot}^{0.24}$.
The activity branches of \cite{BST1998} have been recently supported \citep{BMM17},
but also questioned \citep{RCG17,DLLMS17,BMJRCMPW17,OLKPG17}.
One of the short comings is clearly the use of the
ill-determine convective turnover time $\tau_c$, which is used to the calculated
the Rossby number $\Ro=4\pi P_{\rm rot}/\tau_c$; for every star $\tau_c$ is highly depth-dependent and
different location of a dynamo might invoke a different
$\tau_c$. However, these branches can be also obtained using the
fractional Ca II H\&K emission $R^\prime_{\rm HK}$ instead of the
Rossby number \citep[see e.g.][]{BMM17,OLKPG17}.

Explaining the observational findings via dynamo models has been
challenging. Simple mean-field models of turbulent dynamos produce
rotational dependencies of cycle period similar to the observed ones
using overlapping induction layers \citep{KRS1983}.
Advective dominated flux transport models tend to produce an increase
of cycle periods with increasing rotation rate
\citep[e.g.][]{DC99,BERB02,JBB10}, which is opposite what is
observed. In these models the cycle length is mainly determined by the
strength of return flow of the meridional circulation, which is
believed to decreases with increasing rotation
\citep[e.g.][]{Ko70,BBBMT08,WKKB16,KKOWB17,VWKKOCLB17}.
However, the models of \cite{KR99} show an increase of meridional
circulation strength with rotation, leading to a decrease of cycle
period with rotation \citep[e.g.][]{KRS01}, which agrees qualitatively
with observations.

Another possibility to explain cycles in dynamo models is via the
turbulent (eddy) magnetic diffusivity. In a propagating $\alpha\Omega$ dynamo
wave, the dynamo drivers, whose are responsible for the cycle length,
have to balance with the contribution from the turbulent diffusion.
Using a turbulent diffusivity of
$\etat=2\times10^8\,\rm m^2/s$ one gets a magnetic cycle length of around
23 yrs \citep[e.g.][]{RS72}, which is pretty close to solar value of
22 yrs.
A change in the cycle length can be then associated with a change in
the turbulent diffusion caused by magnetic or rotational quenching
\citep[e.g.][]{RKKS94}. These authors found a cycle dependence of
$P_{\rm cycl} \propto P_{\rm rot}^{0.1}$.

There has been only a limited number of studies of rotational
dependencies of dynamo cycles using global dynamo simulation.
\cite{SBCBN17} found $P_{\rm cycl} \propto P_{\rm rot}^{-1.06}$ in a rather limited sample of rotation rates.
In the recent work by \cite{VWKKOCLB17}, the author found no clear
dependency of cycle periods with rotation. However, a decrease in
cycle period with increasing rotation seems be more likely than an
increase, which might be because of the strong oscillatory
non-axisymmetric magnetic fields in these simulations.

In this work, we present the results of spherical convective dynamo
models with rotation rates varying by a factor of 30. We will determine
the cycle dependency on rotational influence, see
\Sec{sec:magcyc}, interpreted the data in terms of the
Parker-Yoshimura rule by using test-field obtained transport
coefficients, see \Sec{sec:cause}, and compare the finding with
observational as well as other numerical results in \Sec{sec:comp}.

\section{Model and setup}
\label{sec:model}

The detailed description of the general model can be found in \cite{KMCWB13} and will not
be repeated here.
We model the convection zone of a solar-like star in spherical polar
coordinates ($r,\theta,\phi$) using the wedge assumption $0.7\, R<r<R$,
$\Theta_0<\theta<\pi-\Theta_0$ and $0<\phi<\pi/2$ with $R$ being the
stellar radius and $\Theta_0=15^\circ$.
We solve the evolution equations of compressible magnetohydrodynamics
for the magnetic vector potential $\AAA$, which therefore ensures the
solenoidality of the magnetic field $\BB=\nab\times\AAA$, for the
velocity $\uu$, the specific entropy $s$ and density $\rho$. The
model assumes an ideal gas for the equation of state.
The fluid is also influenced by Keplerian gravity and rotation via
the Coriolis force.
Because of the wedge assumption, we use periodic boundary condition in
the azimuthal ($\phi$) direction. 
We assume a stress-free condition for the
velocity field at other boundaries and perfect conducting latitudinal
and bottom boundaries and radial field condition at the top boundary
for the magnetic field.
The energy is transported into the system via a constant heat flux at
bottom boundary and the temperature obeys a black body condition.
On the latitudinal boundaries, the energy flux is vanishing using zero
derivative for the thermal-dynamical quantities.
The detailed setup including the exact equations and expression for
the boundary condition can be found in \cite{KMCWB13, KKOWB17} and
\cite{WKKB14}.

We characterize our runs with the following non-dimensional input parameters;
the Taylor number, the SGS, and magnetic Prandtl numbers
\begin{equation}
\Ta=[2\Omega_0 (0.3R)^2/\nu]^2,\quad \PrSGS={\nu\over\chiSm},\quad \Pm={\nu\over\eta},
\end{equation}
where $\nu$ and $\eta$ are the constant kinematic viscosity and
magnetic diffusivity, and the sub-grid-scale (SGS) heat diffusivity
$\chiSm=\chiS(r_{\rm m})$ is evaluated at $r_{\rm m}=0.85R$.
Furthermore, we use the Rayleigh number obtained from the hydrostatic stratification, 
evolving a 1D model, given by
\begin{eqnarray}
\Ra\!=\!\frac{GM(0.3R)^4}{\nu \chiSm R^2}
  \bigg(-\frac{1}{c_{\rm P}}\frac{{\rm d}s_{\rm hs}}{{\rm d}r}
  \bigg)_{(r=0.85R)},
\label{equ:Ra}
\end{eqnarray}
where $s_{\rm hs}$ is the hydrostatic entropy.
As diagnostic parameters, we quote the density contrast.
\begin{equation}
\Gamma_\rho\equiv\rho(r=0.7R)/\rho(R),
\end{equation}
the fluid and magnetic Reynolds numbers
and the P\'eclet number,
\begin{equation}
\Rey=\frac{\urms}{\nu \kf},\quad \Rm=\frac{\urms}{\eta \kf},\quad
\Pe=\frac{\urms}{\chiSm \kf},
\end{equation}
where $\kf=2\pi/0.3R\approx21/R$ is an estimate of the wavenumber of
the largest eddies.
We defined the Coriolis number as
\begin{equation}
\Co=2\Omega_0\tau_c\equiv\Roo,
\label{eq:Co}
\end{equation}
where $\tau_c=1/\urms \kf$ is the convective turnover time and
$\urms=\sqrt{(3/2)\brac{u_r^2+u_\theta^2}_{r\theta\phi t}}$ is
the rms velocity and the subscripts indicate averaging over $r$,
$\theta$, $\phi$ and a time interval covering the saturated state.
The duration of the saturated state is indicated by $\tau_{\rm sat}$
and covers several magnetic diffusion times.
The kinetic and magnetic energy density are given by
\begin{equation}
E_{\rm kin}={1\over 2}\brac{\rho\uu^2}_{r\theta\phi t}
 ,\quad E_{\rm mag}={1\over 2\mu_0}\brac{\BB^2}_{r\theta\phi t}.
\end{equation}
All values for these non-dimensional input and diagnostic parameters
are shown in \Tab{runs} for all runs.

\begin{table*}[t!]\caption{
Summary of Runs.
}\vspace{12pt}\centerline{\begin{tabular}{l|rrccr|rrrrrc|r}
Run & $\Ot$ &$\Ta$[$10^{6}$]& $\Ra$[$10^{7}$] &$\Rey$&$\Co$&$P_{\rm cycl}$[yr]&$E_{P_{\rm cycl}}$[yr]&$\tilde{P}_{\rm cycl}$[yr]&$\tilde{E}_{P_{\rm cycl}}$[yr]&$P_{\rm PY}$[yr]&$E_{\rm mag}/E_{\rm kin}$&$\tau_{\rm sat}$[yr]\\[.8mm]   
\hline
\hline\\[-2.5mm]
M0.5 &  0.5 & 1.3 &  4.0 & 44 & 0.7   &   1.6 & 1.3 &   3.9 & 12.4 &2.7&0.06&68\\
M1    &  1.0 & 5.4 &  4.0 & 40 & 1.5   & 20.4 & 9.3 & 34.7 & 19.9 &2.1&0.16&104\\%running
M1.5 &  1.5 & 12  &  4.0 & 39 & 2.2   & 23.1 & 9.4 & 46.2 & 32.6 &2.3&0.17&92\\
M2    &  2.0 & 22  &  4.0 & 40 & 2.9   & 10.8 & 2.4 & 33.1 & 23.4 &2.2&0.10&66\\
M2.5 &  2.5 & 34  &  4.0 & 40 & 3.7   & 12.1 & 4.1 &   7.7 & 16.3 &1.8&0.10&61\\
M3    &  3.0 & 49  &  4.0 & 39 & 4.5   &   6.4 & 0.7 &   4.6 & 19.5 &2.1&0.13&64\\
M4    &  4.0 & 86  &  4.0 & 36 & 6.6   &   2.4 & 0.1 &   2.6 &    0.3&1.6&0.21&47\\
M5    &  5.0 & 35  &  4.0 & 34 & 8.6   &   2.2  & 0.1 &   2.3 &   0.3&1.9&0.29&70\\
M7    &  7.0 & 264 &  4.0 & 31 & 13.4 &  2.6  & 0.1 &   2.7 &   0.5&2.4&0.39&51\\
M10  & 10.0& 540 &  4.0 & 27 & 21.5 &  2.7  & 0.2 &   2.5 & 16.9&2.7&0.52&53\\
M15  & 15.0& 1897 &  7.4 & 27 &40.3&  3.8 & 0.2  &   3.8 & 18.8&4.2&0.85&61\\
\hline
\hline
\label{runs}\end{tabular}}\tablefoot{
Second to fourths columns: input parameters.
Last ninth columns: diagnostics computed from the saturated states of
the simulations. $\tau_{\rm sat}$ indicate the time span of the
saturated stage. $P_{\rm cycl}$ and $\tilde{P}_{\rm cycl}$ are the
cycle periods determined from the magnetic field components and the
magnetic energy, respectively, $E_{P_{\rm cycl}}$ and
$\tilde{E}_{P_{\rm cycl}}$ are there corresponding error estimates,
see \Sec{sec:magcyc}. $P_{\rm PY}$ are the cycle periods determined
using a Parker-Yoshimura dynamo wave, see \Eqs{eq:wpy}{eq:Ppy} and
\Sec{sec:cause}.
$E_{\rm mag}/E_{\rm kin}$ is ratio of magnetic to kinetic energy.
All runs have $\PrSGS=2$ and $\Pm=1$ and a density
contrast of $\Gamma_\rho=31$.
}
\end{table*}

The wedge assumption in the azimuthal ($\phi$) direction allows us to
suppress non-axisymmetric dynamo mode with azimuthal degree $m=1,2,3$ and therefore use
the mean-field decomposition to describe the large-scale velocity and
magnetic field.
We use an over-bar to refer to the mean, azimuthal averaged, quantity and a
prime for the fluctuating one, e.g. $\BB=\meanBB+\fluc{\BB}$.

To determine some of the turbulent transport coefficients in these
simulations, we make use of the test-field method
\citep{SRSRC05,SRSRC07,WRTKKB17}.
This method uses linear independent test-fields, which to do not
back-react on the flow to determine the electromotive forces of these
test-fields using the mean and fluctuating flow field of the simulations.
This allows to obtain the all components of the turbulent transport
tensors.
In this work, we will only use the $\phi\phi$ component of the $\alpha$
tensor.

Some of the results, we will present in physical units by using a
normalization based on the solar rotation rate
$\Omega_\odot=2.7\times10^{-6} \s^{-1}$,  the solar
radius $R=7\times10^{8} \m$, the density at the bottom
of the convection zone $\rho(0.7R)=200 \kg/\m^3$, and
$\mu_0=4\pi\cdot10^{-7}$~H~m$^{-1}$.
Furthermore, the rotation of the simulations is given in terms of
solar rotation rate with $\tilde\Omega\equiv\Omega_0/\Omega_\odot$.
However, the rotational influence on the convection is much better
described by the use of the Coriolis number $\Co$.

\section{Results}

For all the simulations we keep all input parameter constant, except
that we increase the rotation rate for $0.5$ to $15$ solar rotation
rate, corresponding to $\Co=0.7$ to $40.3$.
Only for the run with the highest rotation rate (M15), we lower the
diffusivities ($\nu,\eta,\chiSm$) to keep the Reynolds and P\'eclet
numbers on a similar level, however the Prandtl numbers are kept fixed.
We name the runs with 'M' because of their magnetic nature followed by
their solar rotation rate.
Run~M5 has been discussed as Run~I in \cite{WKKB14}, as Run~A1 in
\cite{WKKB16}, as Run~D3 in \cite{KKOWB17}, in \cite{WRTKKB17} and as
Run~G$^{\rm W}$ in \cite{VWKKOCLB17}. 
Run~M3 have been analyzed as Run~B1 in \cite{WKKB16}.
Runs~M10 and M15 are similar to Runs~I$^{\rm W}$ and J$^{\rm W}$ of
\cite{VWKKOCLB17}.
As calculated in \cite{WRTKKB17}, the Rayleigh number for Run~M5 is
around 100 times the critical value. We expect that this factor
increases for lower rotation and decrease for higher rotation.

In this work, we will not discuss all properties of the rotational
influence of the hydrodynamical dynamics, i.e. the angular momentum
evolution, we will instead focus on the discussion and analysis of the dynamo
cycles and their possible origin.

\begin{figure}[t!]
\begin{center}
\includegraphics[width=0.75\columnwidth]{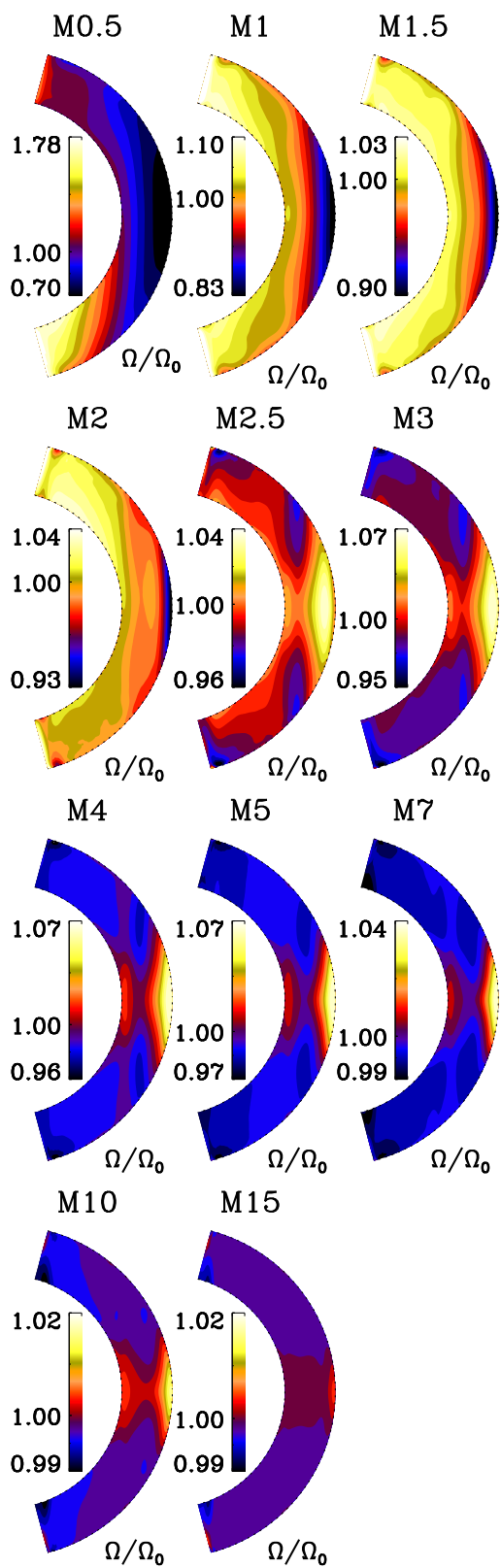}
\end{center}\caption[]{
Normalized differential rotation $\Omega/\Omega_0$ with
$\Omega=\Omega_0+\meanuu/r\sin\theta$ for all
runs. $\Omega$ has been calculated as a time average over the
saturated state.
}\label{diffrot}
\end{figure}

Before we do this in detail, let us look at the differential rotation
generated in these simulation by the interplay of rotation and
turbulent convection.
In \Fig{diffrot}, we show the time averaged differential rotation
$\Omega=\Omega_0+\meanuu/r\sin\theta$ for all runs.
In agreement with earlier findings
\citep{GYMRW14,KMB14,FF14,KKKBOP2015,VWKKOCLB17}, for slow rotation of 
$\Co=0.7$--$2.9$ the equator is rotating slower than the poles,
so-called anti-solar differential rotation and for rapid rotation the
$\Co=3.7$--$40$ the poles are rotating slower than the equator,
so-called solar-like differential rotation.
The overall relative latitudinal and radial shear is the strongest for the
slowest rotation and decreases in the solar-like differential regime
for higher rotation rate. This agrees quantitatively with what is found in
observation \citep{RRB13,LJHKH16}, with mean-field models \citep{KR99}
and previous global simulations
\citep[e.g.][]{KMB11,GYMRW14,KMB14,VWKKOCLB17}.
The difference between the northern and southern hemisphere, for
example in Runs~M0.5 and M2, is caused by a hemispheric dynamo producing
stronger magnetic field in one hemisphere than in the other, see \Sec{sec:magcyc}. 
An interesting feature is the occurrence of a region of a minimum of
$\Omega$ at mid-latitude in the solar-like differential rotation
regime. This region corresponds to area of large negative shear and has
been shown to be responsible for the equatorward migrating dynamo wave
in Run~M5 \citep{WKKB14,WKKB16,KKOWB17,WRTKKB17}.
This region seems to become less pronounced for more rapidly rotating
simulations.

\begin{figure*}[t!]
\begin{center}
\includegraphics[width=2.00\columnwidth]{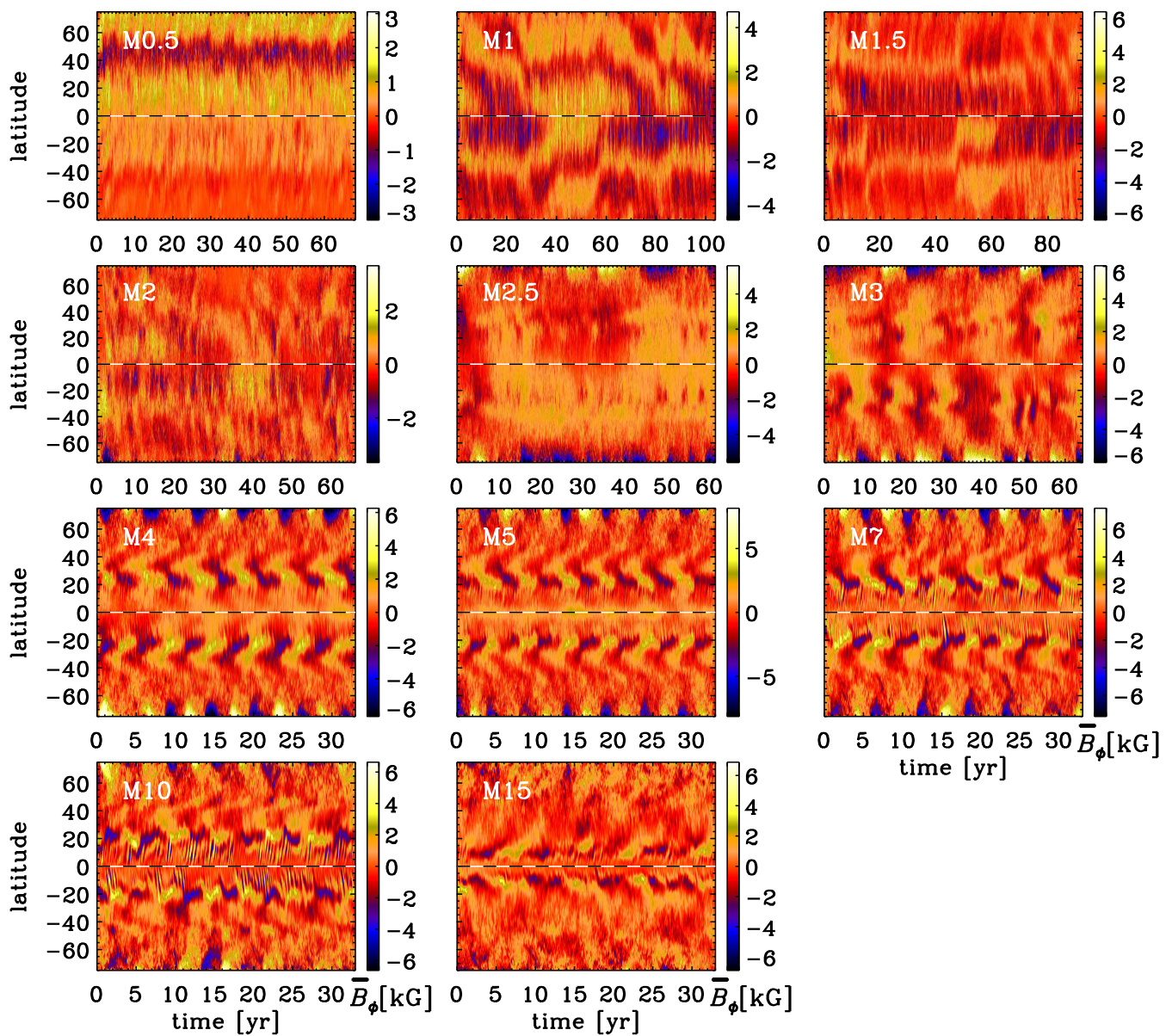}
\end{center}\caption[]{
Mean azimuthal magnetic field $\meanB_\phi$ as a function of time in
years and latitude near the surface ($r=0.98R$) for all runs. The time interval shows the full
duration of the saturated state for Runs~M0.5 to M3 and an interval
of 32 yrs for Runs~M4 to M15 to illustrate the similarity in cycle
length. The black-white dashed horizontal line indicates the equator.
}\label{but}
\end{figure*}

\subsection{Magnetic cycles}
\label{sec:magcyc}

\begin{figure}[t!]
\begin{center}
\includegraphics[width=0.9\columnwidth]{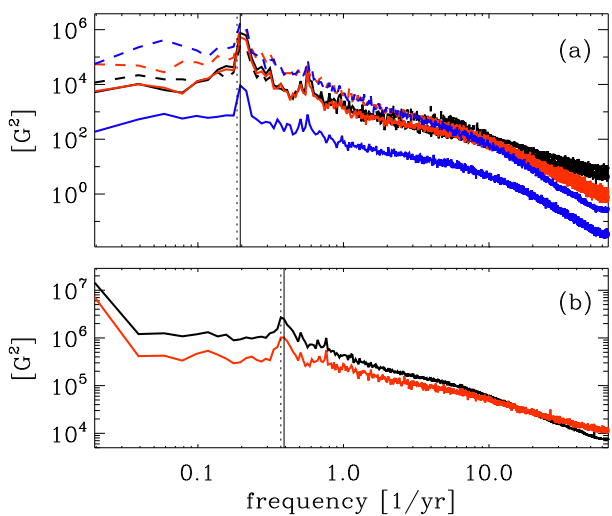}
\end{center}\caption[]{
Magnetic power spectra for Run~M7. (a) Spectra of the mean radial
magnetic field $\meanB_r$ (solid) and the mean azimuthal magnetic
field $\meanB_\phi$ (dashed) near the surface ($r=0.98$; black), in
the middle of the convection zone ($r=0.85$; red) and at the bottom of
the convection zone ($r=0.72$; blue). (b) Spectra of $\meanB{}^{\rm
  rms}=\sqrt{\meanB_r{}^2+\meanB_\theta{}^2+\meanB_\phi{}^2}$ near the
surface ($r=0.85$; red) and averaged over radius (black). All spectra
are obtained for each latitude and then averaged.
The peaks in a correspond to magnetic cycle periods and in b to
activity cycles periods. The solid vertical lines indicate the cycle
periods determine from the weighted average of the spectra in a ($P_{\rm cycl}$), the
dashed lines indicate the cycle periods determine from the weighted
average of the spectra in b ($\tilde{P}_{\rm cycl}$).
}\label{spec}
\end{figure}

\begin{figure}[t!]
\begin{center}
\includegraphics[width=0.9\columnwidth]{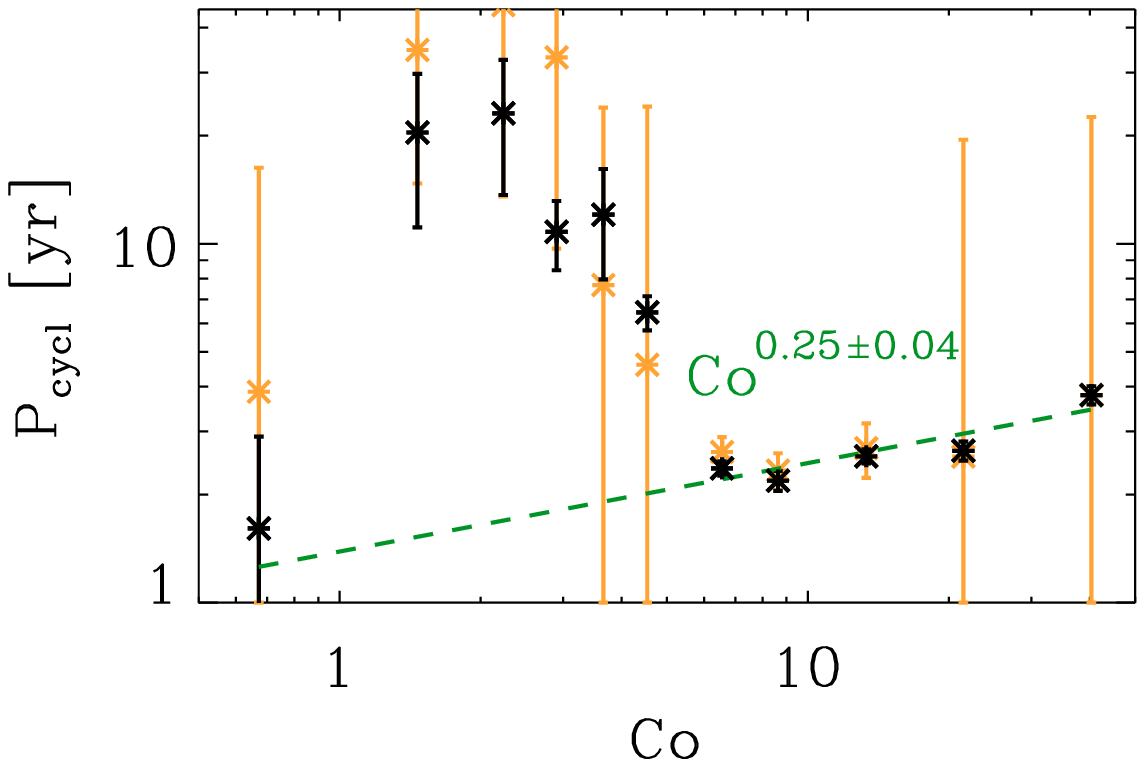}
\end{center}\caption[]{
Cycle periods as a function of Coriolis number $\Co$ showing
$P_{\rm cycl}$ in black and $\tilde{P}_{\rm cycl}$ in orange. The green 
dashed line indicates a power law fit of the Runs~M4 to M15.
}\label{cycle1}
\end{figure}

All simulations discussed here show large-scale dynamo
action. Lowering the rotation rate below $\Ot=0.5$, produces only a weak
large-scale magnetic field ($\Ot=0.4$) or no dynamo action
($\Ot\le0.3$) for the same parameters otherwise.
A small-scale dynamo is not present in any of these simulations
\citep{WRTKKB17}.
In \Fig{but}, we show the near-surface mean azimuthal magnetic field $\meanB_\phi$
as function of time and latitude, so-called butterfly diagram for all
runs.
For the slow rotating Run~M0.5, we find the dynamo produces a
large-scale magnetic field most pronounced in one hemisphere and with
no polarity reversals, only the amplitude shows weak cyclic variations.
For Runs~M1 to M2.5 the magnetic field is of chaotic nature with
polarity reversal, which seems not to be cyclic.
This is similar, what have been found before for slowly rotating
convective dynamos \citep{FF14,KKKBOP2015,HRY16,KKOWB17}.
Interestingly, Run~M1 shows some indication of an oscillating magnetic
field with anti-solar differential rotation, however, from the current
running time, we cannot draw any certain conclusions.
Indication of cyclic solution in the anti-solar regime have been also
found by \cite{KKKBOP2015}, but only recently \cite{VWKKOCLB17} have
obtained clear cyclic solutions with many polarity cycles. 
Run~M3 shows indication of a cyclic magnetic field, most pronounced at
the poles. The magnetic cycle is even more pronounced in the middle of
convection zone, as shown in \cite{WKKB16}.
The Runs~M4 to M15 show a clear cyclic magnetic field with regular
polarity reversals. The cycle length for these runs seems to be very similar.
For Runs~M4 to M10 the magnetic field shows a clear equatorward
propagation similar what it observed for the solar activity belt. 
Furthermore, a shorter, much weaker poleward migrating cycle is present in addition to
the equatorward migrating mode. It seems to become stronger for
increasing rotation. This weak short cycle has been associated with a
local $\alpha^2$ dynamo mode in addition to the $\alpha\Omega$ dynamo
causing the equatorward migration \citep{KKOBWKP16, WRTKKB17}.
For Run~M15 the field shows indication of both equatorward and poleward
migration.

To quantify the cycle period of the magnetic field of all runs, we
calculate the power spectrum of the magnetic field and use the
strongest peak as the cycle frequency.
In the following we distinguish between magnetic cycle and activity
cycle. The magnetic cycle includes a full polarity reversal,
corresponding to the 22 yrs on the Sun, the activity cycle uses the
maximum and minima of the magnetic energy, so 11 yrs for the Sun.
We use two ways to calculate our cycle period, for the first we
determine the magnetic cycle and take the half and for the second we
determine the activity cycle, both results are shown in \Tab{runs}.
For the magnetic cycle we take the radial and azimuthal mean magnetic
field component and calculating the power spectrum for each latitude
and at three radii ($r=0.98, 0.85, 72$). Then we average the spectra
over latitude to obtain six spectra. Now, we determine the frequency of
largest speak of each spectrum with a corresponding error. 
The error is estimated by taking the full-width-half-maximum unless
its narrower than the local grid spacing in the frequency space, in
which case we take local grid spacing as an error estimate.
From the six frequencies, we calculate the weighted average of the
corresponding periods. We take the half of the averaged magnetic cycle
period and
show it as $P_{\rm cycl}$ with its error $E_{P_{\rm cycl}}$ in
\Tab{runs}.

For the activity cycle, we use the root-mean-squared value of the
magnetic field $\meanB{}^{\rm
  rms}=\sqrt{\meanB_r{}^2+\meanB_\theta{}^2+\meanB_\phi{}^2}$ near the
surface and as a function of radius. As above we calculate the
spectrum for each latitude and radius and averaged over them.
The cycle period is determined in the same way as for magnetic cycle,
without dividing by two. The results are shown as
$\tilde{P}_{\rm cycl}$ with its error $\tilde{E}_{P_{\rm cycl}}$ in \Tab{runs}.
In \Fig{spec}, we show as an example the power spectra of
Run~M7. Beside the cycle with an activity period of around 2.7 yr, we notice the weak short cycle,
which is also visible in \Fig{but}.

For the Runs~M4 to M15, the cycle periods can be determined well with a
small error in $P_{\rm cycl}$. Furthermore, for these Runs
$\tilde{P}_{\rm cycl}$ agrees very well with $P_{\rm cycl}$, even
though their errors are higher. The larger errors of $\tilde{P}_{\rm
  cycl}$ are caused by the summation over phase-shifted magnetic field
components and this can result in a less pronounced peak in the spectrum of
$\meanB{}^{\rm rms}$.
This can be also seen in the fact that $\tilde{E}_{P_{\rm cycl}}$ is
for all runs significant larger than $E_{P_{\rm cycl}}$.
For the slowly rotating runs, the periods are not well determined with
significant differences with between the two methods and also large errors.
In the following, we therefore focus on the analysis of the runs with a
clear cycle period.

In \Fig{cycle1} we show the cycle periods as a function of Coriolis
number.
We find two clearly separated group of runs: the slowly rotating
simulations with long ill-determined cycles (Runs~M1 to M3) and the
moderately to rapidly rotating runs with short cycles (Runs~M4 to
M15).
In the latter group the cycle periods increase slightly with
increasing rotation. We perform a power law fit for theses runs and
obtain $P_{\rm cycl} \propto \Co^{0.25\pm 0.04}$, or in terms of rotation
period $P_{\rm cycl} \propto P_{\rm rot}^{-0.33\pm 0.05}$.
This value is in disagreement with the observation of
\cite{Noyes+al:1984a} and \cite{OKPSBKV16}, but agree qualitatively
with scaling of advective dominated flux-transport dynamo models
\citep[e.g.][]{DC99,BERB02,JBB10}.
\cite{SBCBN17} also found an increase of cycle period with increasing
rotation, however their power law fit reveals a much steeper increase
with rotation $P_{\rm cycl} \propto P_{\rm rot}^{-1.06}$.
The cycle period calculated for Run~M5 agrees with the cycle periods
obtain for similar runs using the D2 phase dispersion statistics and
the Ensemble Empirical Mode Decomposition \citep{KKOBWKP16,KKOWB17}.

\subsection{Cause of magnetic cycles}
\label{sec:cause}

\begin{figure}[t!]
\begin{center}
\includegraphics[width=0.32\columnwidth]{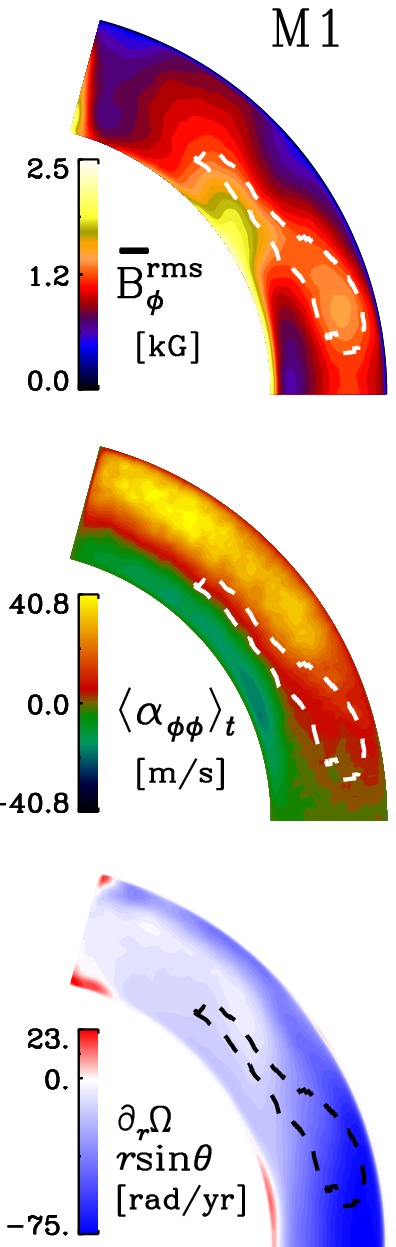}
\includegraphics[width=0.32\columnwidth]{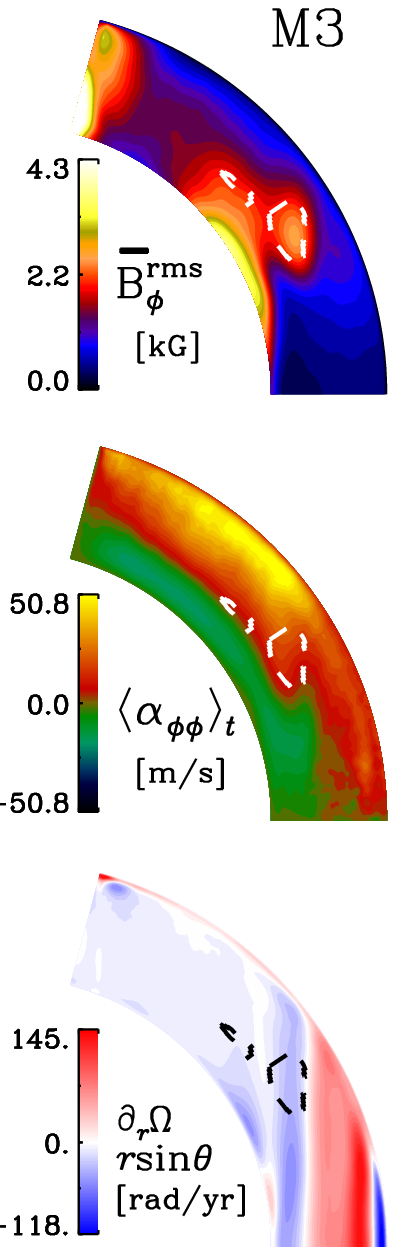}
\includegraphics[width=0.32\columnwidth]{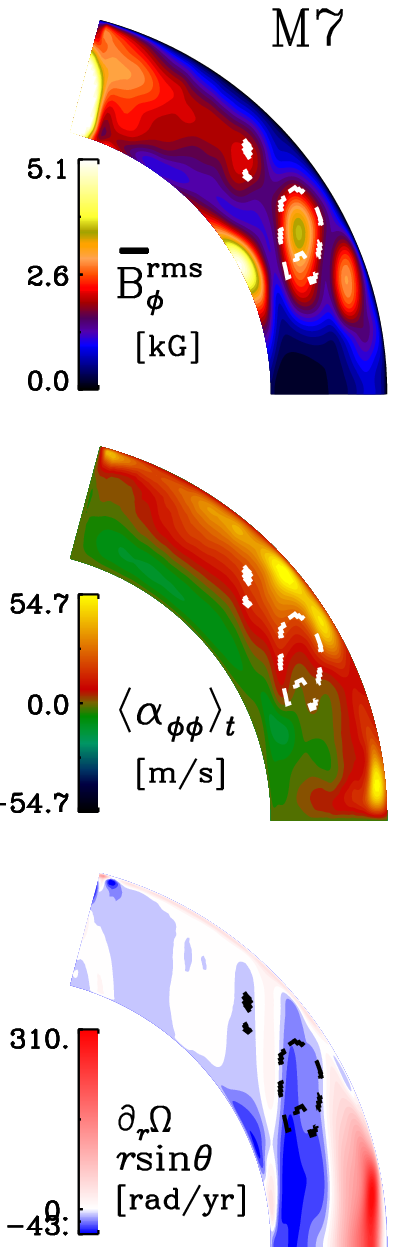}
\end{center}\caption[]{
Mean magnetic field, $\alpha$ effect and radial shear profiles for
Runs~M1, M3 and M7.
We show the rms mean azimuthal magnetic field averaged over the
saturate state $\meanB{}_\phi^{\rm rms}$ (top row), $\alpha_{\phi\phi}$
determined with the test-field method (middle row) and the radial shear
$r\sin\theta \partial\Omega/\partial r$ (bottom row).
The dashed lines indicate the region where we calculate $P_{\rm PY}$.
}\label{PY2}
\end{figure}

Earlier studies of similar simulations as Run~M5 show that the
equatorward migrating mean magnetic field can be well explained with a
Parker-Yoshimura \citep{P55,Yos75} $\alpha\Omega$-dynamo wave
propagating equatorward \citep{WKKB14,WKKB16, WRTKKB17}.
Following the calculation of \cite{P55} and \cite{Yos75}, we can
compute the cycle frequency of the dynamo wave using \citep[see also][]{Stix1976}
\begin{equation}
\omega_{\rm PY}=\left|\frac{\alphapp\,k_\theta}{2}r\cos\theta\frac{\partial\Omega}{\partial  r}\right|^{1/2},
\label{eq:wpy}
\end{equation}
where $k_\theta$ is the latitudinal wave number.
The corresponding activity cycle period is then given by 
\begin{equation}
P_{\rm PY}={2\pi\over 2\,\omega_{\rm PY}}.
\label{eq:Ppy}
\end{equation}
As pointed out by \cite{WKKB14}, to get a meaningful
result for the direction and therefore the period of dynamo wave, the
location where measuring the shear and the $\alphapp$ is crucial.
Following this work, we calculate $P_{\rm PY}$ in the region
where i) $\meanB{}_\phi^{\rm rms}=(\bra{\meanB{}_\phi^2}_t)^{1/2}$ is large, in our case at least
larger than the half of the maximum value, ii) the radial shear
$\partial\Omega/\partial  r$ is
negative and iii) $\alphapp$ is positive.
The last two criteria are needed to excite an equatorward migrating
dynamo wave, following the Parker-Yoshimura sign rule.
To make sure that these drivers are really responsible to excite a dynamo
wave at this location, the production of azimuthal magnetic field must be
large at this location, leading to criterion i). The lower limit of
half of the maximum value is a reasonable choice; a slightly different
value has only little effect on the cycle period determination  and
on its dependence on rotation rate.
The criteria have been also successfully used to confirm
Paker-Yoshimura dynamo waves in similar simulations
\citep{WKKB14,WKKB16,WRTKKB17,KKOBWKP16}.
Using these criteria, we then average over these regions.
In \Fig{PY2}, we show these regions for Run~M1, M3 and M7.
To calculate $P_{\rm PY}$, we choose 
\begin{equation}
k_\theta={1\over R (1 -2\Theta_0/\pi),} = 1.2/R,
\end{equation}
where the factor $1 -2\Theta_0/\pi$ takes into account the absent of the poles in
our simulations.
However, the actually value of $k_\theta$ will only affect the value
$P_{\rm PY}$ with a $-1/2$ dependency, but not the scaling with
rotation.

\begin{figure}[t!]
\begin{center}
\includegraphics[width=0.9\columnwidth]{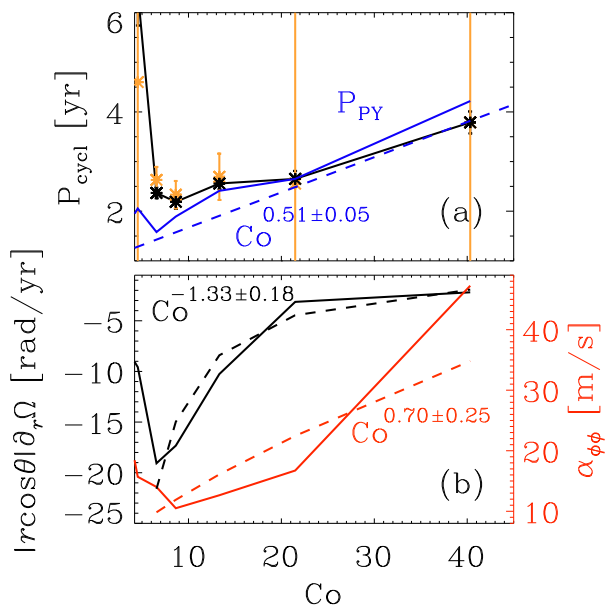}
\end{center}\caption[]{
(a) Comparison of the cycle periods $P_{\rm cycl}$ (black) and
$\tilde{P}_{\rm cycl}$  (orange) with predicted cycle
periods using a Parker-Yoshimura dynamo wave $P_{\rm PY}$ (blue) for
Runs~M4 to M15.
The dashed blue line indicates a power law fit to $P_{\rm PY}$.
(b) Contributions to the Parker-Yoshimura dynamo wave containing the
radial shear (black line; left y-axis) and $\alpha_{\phi\phi}$ (red;
right y-axis) for Runs~M4 to M15. The dashed lines indicate the
corresponding power law fits.
}\label{PY}
\end{figure}

In \Tab{runs}, we list all computed values for $P_{\rm PY}$ in the
eleventh column and they agree well with the values of $P_{\rm cycl}$ and
$\tilde{P}_{\rm cycl}$ for the runs with well determined cycles (M4 to
M15).
For oscillatory solution of planetary dynamos, \cite{GDW12} found also
a good agreement between rotational dependency of measured and dynamo wave predicted cycle
length.
In \Fig{PY}a, we show for these runs the cycle periods $P_{\rm cycl}$
and $\tilde{P}_{\rm cycl}$ together with the predicted period $P_{\rm
  PY}$.
A power-law fit results in $P_{\rm PY}\propto\Co^{0.51\pm 0.05}$ which is close to $P_{\rm cycl} \propto
\Co^{0.25\pm 0.04}$.
Therefore, the Parker-Yoshimura dynamo wave explains well the weakly
dependency of cycle frequency with rotation, we find for the
moderately and rapidly rotating simulations.
We now go a step further and check, which mechanism of dynamo wave
causes this rotational dependency.
For this we plot in \Fig{PY}b the rotational dependency of the radial
shear and $\alpha$ effect in terms of
$|r\cos\theta|\partial\Omega/\partial r$ and $\alphapp$; as for
$P_{\rm PY}$ both quantities are averaged over the region of
interest.
The strength of the shear strongly weakens for larger rotation, with an
estimated scaling of $\Co^{-1.33\pm 0.18}$. For Run~M15 the shear in the
region is just below zero explaining the mixture of equatorward and
poleward migration pattern shown in \Fig{but}.
For $\alphapp$, we find an increase with rotation corresponding to a scaling of
$\Co^{0.70\pm 0.25}$, so much less than linear.
The strong decrease in shear causes the cycles to become
larger with rotation: assuming a constant $\alphapp$, shear alone
would leading a scaling of $P_{\rm cycl} \propto \Co^{0.67\pm 0.09}$.
The $\alpha$-effect on the other hand lead to a decrease of cycle
length with rotation; $P_{\rm cycl} \propto \Co^{-0.35\pm 0.12}$.
Because the increase of cycle length due to shear is stronger
than the decrease due to the $\alpha$ effect, the resulting cycle
length shows only a weak increase with rotation.

\begin{figure}[t!]
\begin{center}
\includegraphics[width=0.9\columnwidth]{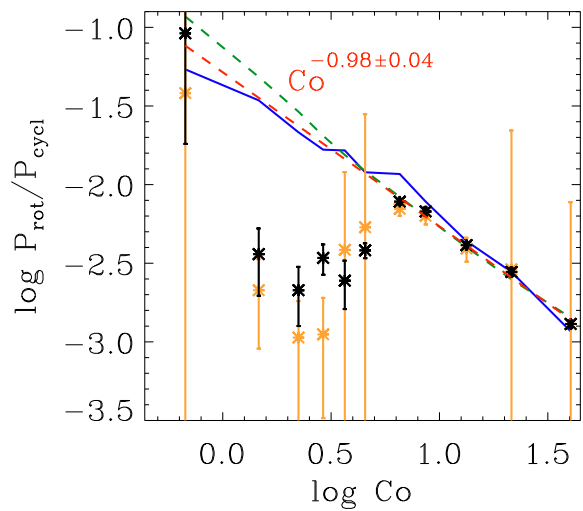}
\end{center}\caption[]{
Ratio of rotation period and cycle period $P_{\rm rot}/P_{\rm cycl}$
over the Coriolis number $\Co$.
The black asterisks indicate $P_{\rm cycl}$ and the orange ones
$\tilde{P}_{\rm cycl}$. The blue solid line shows the predicted cycle
length from \Eqs{eq:wpy}{eq:Ppy}, the green dashed line the fit of
\Fig{cycle1} and the red dashed line a power law fit of Runs~M4 to
M15.
}\label{Prot}
\end{figure}

The surprising issue with interpretation of the magnetic field
evolution as a Parker-Yoshimura dynamo wave is that for runs rotating
slower than the M4 ($\Co=6.5$) it fails.
Equation (\ref{eq:wpy}) for these runs predict cycle periods of similar
length as for the more rapidly rotating runs, but the actual magnetic
field shows no clear cyclic evolution.
For example, Run~M3 shows similar condition for a dynamo wave as in
Run~M7; there exists a localized region, where the mean toroidal field is
strong, $\alphapp$ is positive and shear negative, see \Fig{PY2}.
In the simulations with anti-solar differential rotation
(Run~M0.5 to M2), we find instead a more extended region of strong
mean toroidal field, positive $\alphapp$ and negative shear, however
strength of the shear and the $\alpha$ effect should be sufficient to
excite an $\alpha\Omega$ dynamo wave.
One of the reasons for the absent of an $\alpha\Omega$ dynamo wave can
be the larger turbulent magnetic diffusion due
to higher convective velocities as shown in \Fig{etat}. This is in
agreement with previous studies of rotating convection in Cartesian
boxes \citep{KKB09}.
To have a reliable statement, if a $\alpha\Omega$ dynamo is actually
operating in these simulation, and what is the reason for not
exciting dynamo wave, need to be studied in more detail using all the turbulent
transport coefficients. We postpone such a study to the future.

We can now also interpret the scaling of the shear and the $\alpha$ effect in terms
of mean-field models \citep[e.g.][]{KR80,R89}. From models of
differential rotation, one typically finds that the absolute radial
and latitudinal differential rotation stays nearly constant for
increasing rotation \citep[e.g.][]{KR99}, which disagrees with our findings.
However, we stress here that in these models they take the latitudinal
averaged values at bottom and at surface to compute the radial
differential rotation, we compute the local radial shear in the region
of interest.
In mean-field dynamo models $\alphapp$ is related to the mean
kinetic helicity $\overline{\fluc{\uu}\nab\times\fluc{\uu}}$ and
therefore is linear related to the $\Omega$. Taking also the convective turnover
time $\tau_c$ into account leads to scaling of $\alphapp\propto\Co$
\citep[e.g.][]{KR80}. As shown in \cite{WRTKKB17}, approximating the diagonal
$\alpha$ components with $\alpha_K=
-1/3\tau_c\,\overline{\fluc{\uu} \nab\times\fluc{\uu}}$ is not correct and can
lead to the overestimation of $\alphapp$.
Indeed, we find $\alphapp$ in the region of interest depends weaker
on rotation as predicted from mean-field models. The overall scaling
of $\alphapp$ averaged of the simulations might be different, but the
import value of $\alphapp$ determining the cycle period comes from
this region.
\cite{WRTKKB17} took also into account the non-linear quenching of the
$\alpha$ effect
due to magnetic helicity conservation \citep[see][for details]{BS05} and use
the form introduced by \cite{PFL76} $\alpha=
-1/3\tau_c\,\overline{\fluc{\uu}\cdot\nab\times\fluc{\uu}} + 1/3
\tau_c/\rho\,\overline{\nab\times\fluc{\bb}\cdot\fluc{\bb}}$, but
still could not find a agreement with the actual measured $\alphapp$,
see Fig. 1 and 2 of \cite{WRTKKB17}.

\begin{figure}[t!]
\begin{center}
\includegraphics[width=0.9\columnwidth]{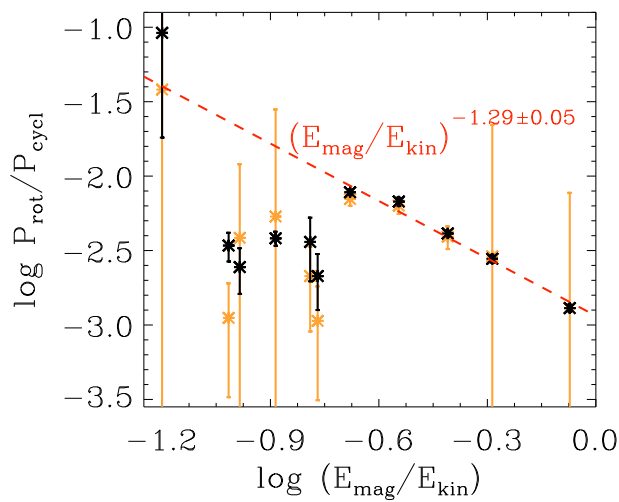}
\end{center}\caption[]{
Ratio of rotation period and cycle period $P_{\rm rot}/P_{\rm cycl}$
over magnetic and kinetic energy ratio.
The black asterisks indicate $P_{\rm cycl}$ and the orange ones
$\tilde{P}_{\rm cycl}$.The red dashed line indicates a power law fit of Run~s M4
to M15.
}\label{Prot2}
\end{figure}

\begin{figure}[t!]
\begin{center}
\includegraphics[width=0.9\columnwidth]{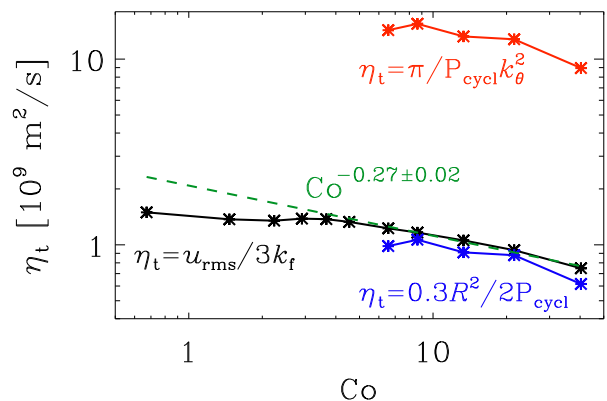}
\end{center}\caption[]{
Turbulent (eddy) magnetic diffusivity over Coriolis number Co.
The diffusivity is determined using an estimate of the turbulence in
the simulations $\etat=1/3\urms/\kf$ (black) and using the cycle
periods as in \Eq{eq:etat1} (red) and as in \Eq{eq:etat2} following
\cite{RS72} (blue). The green dashed line indicates a power law fit of
the black asterisk of Runs~M4 to M15.
}\label{etat}
\end{figure}

Furthermore, we plot the ratio of rotation period and cycle
period $P_{\rm rot}/P_{\rm cycl}$ over Coriolis number, see \Fig{Prot}. We
find a scaling of $P_{\rm rot}/P_{\rm cycl} \propto \Co^{-0.98\pm 0.04}$ for the
runs with well determined cycle. This scaling fits well with the cycle
period predicted by a dynamo wave.
Interestingly, Run~M0.5 fit well to this relation, even though we find
no polarity reversals there.
In interpretation of stellar observation, $P_{\rm rot}/P_{\rm cycl}$
is often used to determine the quenching of the $\alpha$ effect.
If one assumes a linear dependency of $\alpha$ and $\partial
\Omega/\partial r$ on $\Omega$ together with \Eq{eq:wpy}, $P_{\rm rot}/P_{\rm
  cycl}$ over $\Co$ give an estimate over the rotational quenching of
$\alpha$ \citep[e.g.][]{BST1998}.
Moreover, one can go a step further and plot $P_{\rm rot}/P_{\rm
  cycl}$ over magnetic activity, which is related to the surface magnetic
field strengths \citep[e.g.][]{SCZS89}. In case of dynamo simulations
we can instead use the ratio of magnetic and kinetic energy, so-called
dynamo efficiency, to mimic the magnetic activity as done in \Fig{Prot2}.
Then the $P_{\rm rot}/P_{\rm cycl}$ dependence on magnetic activity can be
interpreted as the magnetic quenching of the $\alpha$ effect.
However, doing so stellar observations indicate an increase of the
$\alpha$ effect with magnetic activity in the inactive and active
branch \citep{BST1998,SB1999,BMM17}.

In our simulations, the situation is different. As described above, the
radial shear decreases and the $\alpha$ effect increases with higher rotation rate.
Therefore, we cannot use $P_{\rm rot}/P_{\rm cycl}$ to estimate the quenching
of the $\alpha$ effect.
The scaling of $P_{\rm rot}/P_{\rm cycl} \propto \Co^{-0.99\pm 0.04}$ as shown
in \Fig{Prot} might
seem to be expected, because we plot rotation rate over rotation
rate, but the Coriolis number includes also the strength of
convection, in terms of $\tau_c$, which is influenced by rotation as well.
We do not find any indication of branches with positive slopes similar
to the inactive or active branch as postulated by \cite{BST1998}. 
Furthermore, also the slope is different from what is found from the
super-active branch, which has a slope of $\Co^{-0.43}$
\citep{SB1999}.

In \Fig{Prot2}, we plot $P_{\rm rot}/P_{\rm cycl}$ over ratio of magnetic and
kinetic energy, which can be interpreted as the dynamo efficiency, we
find a scaling of $P_{\rm rot}/P_{\rm cycl} \propto (E_{
\rm mag}/E_{\rm kin})^{-1.29\pm 0.05}$.  
The scaling seems to agree qualitatively with what is found in \cite{VWKKOCLB17}.
Also, here, we do not find any
indication of a positive slope and therefore a similarity to the
activity branches. 
It seems clear the runs with well determined cycles cannot be
interpreted in terms of activity branches with positive slopes.
However, our scaling agrees qualitatively with the suggested
transitional branch by \cite{DLLMS17}.

Another way to analyze the scaling of the cycle period with rotation
is via the turbulent eddy magnetic diffusivity.
In the saturated state the dynamo drivers has to balance to
contribution of the magnetic diffusion.
In a $\alpha\Omega$ dynamo wave, the balance reads
\begin{equation}
\omega_{\rm PY} - k^2\etat =0,
\label{eq:wpy_etat}
\end{equation}
where $k$ is a wavenumber and $\omega_{\rm PY}$ is given by
\Eq{eq:wpy}.
Therefore, we can use this equation to calculate $\etat$ based on the
cycle frequency and compare with estimated values using the turbulent flow
of the simulations.
Using the first-oder-smoothing-approximation \citep[FOSA, see e.g.][]{KR80} and isotropic and
homogeneous turbulence, the turbulent eddy diffusivity can be
estimated as
\begin{equation}
\etat={1\over 3}{\urms\over \kf}.
\label{eq:etat}
\end{equation}
If we assume $k=k_\theta$ in \Eq{eq:wpy_etat}, we can calculate
$\etat$ based on the cycle period
\begin{equation}
\etat={2\pi\over 2\,P_{\rm cycl}\,k_\theta^2}.
\label{eq:etat1}
\end{equation}
Another way is to use the radius $R$ and the depth of the convection
zone $0.3R$ of the star to relate $\etat$ with the cycle period \citep{RS72}.
\begin{equation}
\etat={0.3R^2\over 2\,P_{\rm cycl}}.
\label{eq:etat2}
\end{equation}
We show the values for both expressions for the runs with well determine
cycles (Runs~M4 to M15) together with values of \Eq{eq:etat} as a
function of Coriolis number in \Fig{etat}.
$\etat$ of \Eq{eq:etat2} fits remarkable well with \Eq{eq:etat}.
Even the scaling of $\etat=\Co^{-0.27\pm 0.02}$ determined from \Eq{eq:etat}
is the same as expected from the cycle periods
$\etat\propto 1/P_{\rm cycl}\propto\Co^{-0.25\pm 0.04}$, see \Fig{cycle1}.
This good agreement is indeed interesting and not fully expected, because the estimation of
$\etat$ in \Eq{eq:etat} is based on strong assumptions, which are most
likely not fulfilled in these simulations. 
The values obtained through \Eq{eq:etat1} have obviously the same
scaling as \Eq{eq:etat2}, however the values are around a factor of 10
higher. This is because of the different values of scales/wave numbers
included in this calculation.
If we use $k_\theta$ instead of $\kf$ in \Eq{eq:etat}, the curves
would lie closer together.
Therefore, the scaling of cycle period with rotation rate of $P_{\rm
  cycl} \propto \Co^{0.25\pm 0.04}$ can also be well explained with the
rotational quenching of the turbulent (eddy) magnetic diffusivity.

\begin{figure*}[t!]
\begin{center}
\includegraphics[width=1.9\columnwidth]{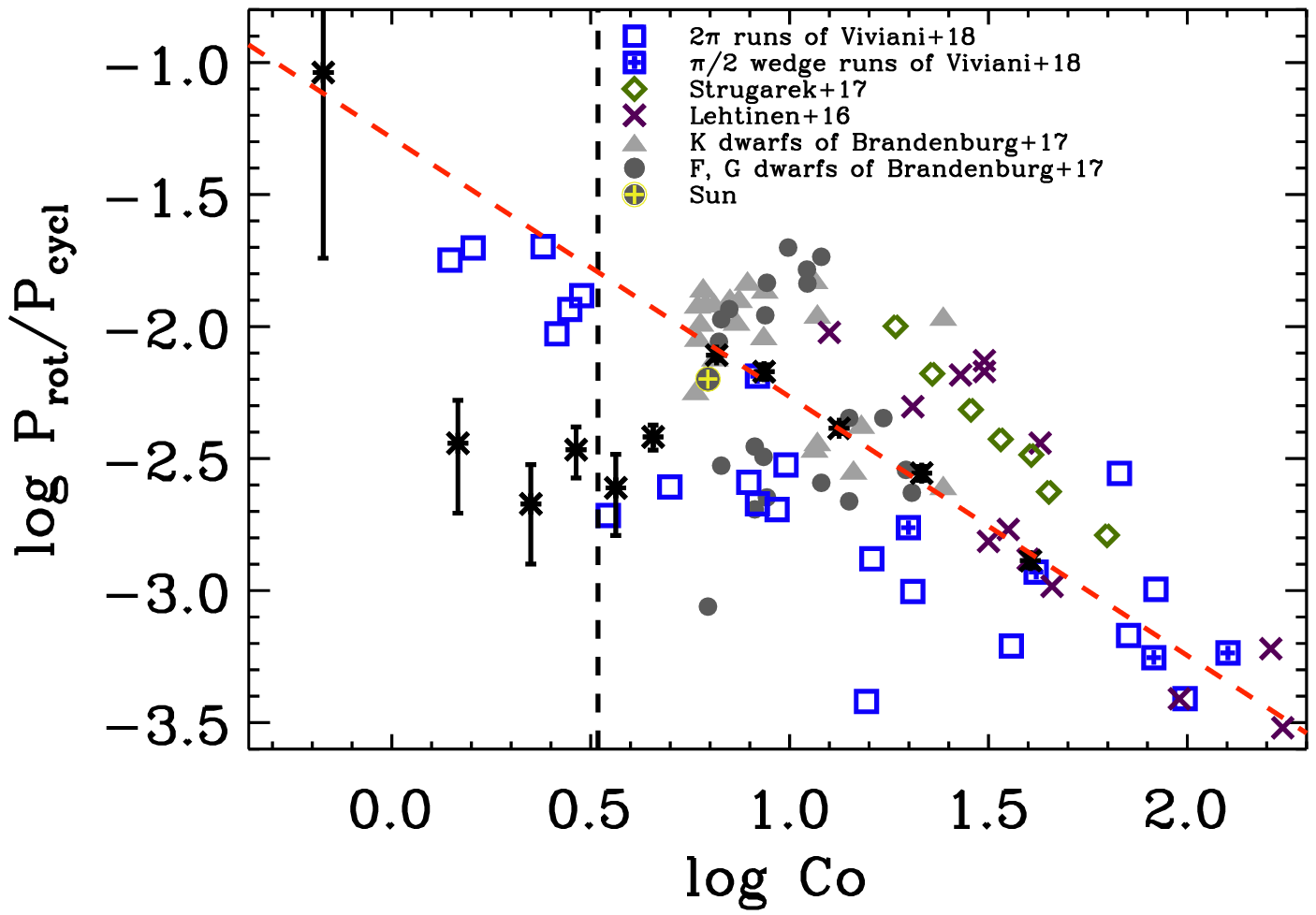}
\end{center}
\caption[]{
Ratio of rotation period and cycle period $P_{\rm rot}/P_{\rm cycl}$
over the Coriolis number $\Co$.
The black asterisks indicate $P_{\rm cycl}$. The dashed red line indicates
the fit $\Co^{-0.98}$ of \Fig{Prot} and the dashed black line the transition from
anti-solar to solar like differential rotation. We include the simulations of \cite{VWKKOCLB17}
(blue squares, with crosses for wedge runs) and \cite{SBCBN17} (green diamond), the
observational studies of \cite{LJHKH16} (purple crosses) and of
\cite{BMM17} (light grey triangles for K dwarfs, dark grey circles for
F, G dwarfs, including the Sun: yellow cross).
}\label{Prot3}
\end{figure*}

\subsection{Comparison with observational and other numerical studies}
\label{sec:comp}

There is only limited amount of numerical studies of cycles of solar
and stellar dynamos. In the following, we compare our results with the
recent work of \cite{SBCBN17} and \cite{VWKKOCLB17}.
The study by \cite{SBCBN17} include seven models, all showing cyclic
dynamo solutions. Even though the authors interpret their models in
the vicinity of the Sun, their Rossby numbers indicate a rapid
rotational regime. If we convert their numbers to our definition of
Coriolis numbers or inverse Rossby number as defined in \cite{BST1998},
respectively,  we obtain values of $\Co=18$ to $62$.
These high numbers are due to their low convective
velocities compared to other models \citep{KKOWB17}.
For comparison the estimated Coriolis number of the Sun is $\Co=6.2$
\citep[e.g.][]{BMM17}.
In \Fig{Prot3}, we plot the models of \cite{SBCBN17} together with our models. There
seems to be no overlap between their and our simulations.
However, also their simulations show a decrease of $P_{\rm rot}/P_{\rm
  cycl}$ with rotation following an even steeper slope.
The similar slope might be caused by a similar dynamo mechanism and
the small difference and the offset might be because of different
system parameters, as Prandtl numbers and/or Reynolds numbers.
As shown in \cite{KKOWB17}, the Reynolds numbers in
typical simulations with the EULAC code can be lower compared to
models of other, similar codes.
This might also explain the dominantly axisymmetric large-scale magnetic field
solution in \cite{SBCBN17}. 
The simulations of \cite{VWKKOCLB17} show clearly that if there is a
transition from axisymmetric to non-axisymmetric magnetic solution at
$\Co\ge 3$. However, if the resolution, therefore the Reynolds and
Rayleigh  numbers, are not high enough, the non-axisymmetric magnetic solution can be not
obtained. This is most important for large rotation rates, where the
convection is rotationally quenched.
Furthermore, \cite{SBCBN17} claim that their dynamo cycle is caused by
a non-linear feedback of the torsional oscillation on the magnetic field. Such an
interpretation is not very likely to be correct, because to have a cyclic torsional
oscillation one needs a cyclic dynamo in the first place.
In the light of the results of this work we are inclined to think that
also their cyclic magnetic field is caused by a Parker-Yoshimura
dynamo wave. Indeed, their differential rotation profiles show
localized regions of strong negative shear, where also the mean magnetic
field propagates equatorward.

The study of \cite{VWKKOCLB17} probe a large range of rotation rate in
particular in the rapid rotational regime. Their simulations using a
similar setup than the one in this work, but for most of their
simulations they use a full $2\pi$ extend in the azimuthal direction
($2\pi$ runs) and obtain non-axisymmetric large-scale field solutions for moderately to rapidly
rotating runs. In \Fig{Prot3}, we also over-plot their simulations.
Their $\pi/2$ wedge runs agrees well with our runs and our obtained scaling of
$\Co^{-0.98\pm 0.04}$. This also means that their estimates of cycle periods
based on the temporal variation in the large-scale magnetic energy
seems to describe the magnetic cycle well.
Their $2\pi$ runs with rapid rotation ($\Co\ge 3$)
show clearly a different scaling, similar to $\Co^{-0.43}$ of the
super-active branch \citep{SB1999,VWKKOCLB17}.
This might mean that the non-axisymmetric magnetic field solutions have
a different scaling than the pure axisymmetric ones.
The argument of different scaling is also supported from the fact that
the slowly rotating axisymmetric simulations of \cite{VWKKOCLB17} can be well described
by the cycle period scaling estimated in this work, see \Fig{Prot3}.
However, surprising is the fact that for nearly the same rotation rate
the $2\pi$ runs show much shorter and much clearer cycles than their
corresponding $\pi/2$ wedge simulations.

At the end, we compare our results with observational obtained stellar
cycles. One sample comes from \cite{LJHKH16}, where the authors use
photometry to measure cyclic variation is solar-like stars. From this
sample, we only plot the cycles which are identified as better than
``poor'' and plot them in \Fig{Prot3}.
For rapid rotating stars, their cycles fall surprising well on our
scaling relation even though their magnetic field is non-axisymmetric.
We note here that \cite{LJHKH16} did not do any direct measurement of
the magnetic field, but inferred the degree of non-axisymmetry from
the spot distributions.
For slowly rotation, the stars seems to fall on a parallel line with a
similar scaling.
\cite{LJHKH16} found that the crossover from the transitional branch
to the super-active branch happens at around $\Co=26.3$ and
chromospheric activity value of $\log R^\prime_{\rm HK} = -4.4$. 
We further include the sub-sample of stars from the Mount Wilson
sample analyzed by \cite{BMM17}.
Our scaling relation falls through these stars; however, a lot of stars
are not captured by our scaling.
As shown by \cite{BMM17}, the stars around $\Co=10$ form the inactive
branch with a positive slope and the stars with higher rotation form
the active branch also with a positive slope. The active branch is not
as confined as the  inactive one. In recent works the existence of
these branches have been questioned \citep{RCG17,DLLMS17}.
Two new studies of the Mount Wilson sample data
\citep{BMJRCMPW17,OLKPG17} only find an indication for an
inactive branch and otherwise a distribution similar to our scaling
showing indication of an transitional branch.
There, the authors re-analyses the full Mount Wilson sample, without
relying of the cycle determination of \cite{Baliunas1995}.
The problem with determining cycles from chromospheric and
photospheric activity time series is due to the method to for the
period search. For star in the inactive branch the cycles are clean
and can be well determined, but more active stars with higher rotation
rates exhibit a complex behavior with multiple cycles
\citep[e.g.][]{OKPSBKV16}.
There, the calculated cycle period can differ depending with method is
used \citep[e.g.][]{OLKPG17}.
We also note that one to one comparison with observation is not always
possible. Even though using the Coriolis number is more meaningful than
then the rotational period, because it indicates if the star or
simulation is in the slow or rapidly rotating regime, the convective
turnover time is usually ill determined for observed stars.
Furthermore, as we know the turnover time and therefore
the corresponding Coriolis number in the Sun changes several orders of
magnitude from the solar surface to the bottom of convection zone
\citep[e.g.][]{Stix:02}, it is difficult even for the Sun to
estimate a single number as a meaningful Coriolis number.
Therefore, \cite{BMM17} and \cite{OLKPG17} use the chromospheric
activity instead of the Coriolis number for their analysis.
Using this, \cite{OLKPG17} finds that the cycle periods of this work
fit very well with their observed stellar cycles, see their Fig. 6.

Interestingly the cycle data taking from \cite{BMM17} indicate that
the Sun lies close to our scaling relation, actually close to Run~M4.
If we go a step further and assume the Sun's 22 yrs magnetic cycle is
caused by a Parker-Yoshimura dynamo wave with a cycle-rotation scaling
similar to our simulations, we can estimate the
corresponding Coriolis number and then with the definition of \Eq{eq:Co} we
also calculate the corresponding value of $\urms$.
We calculate a Coriolis number of $\Co=8.5$ corresponding to $\urms=21.5 \mpsec$. This velocity
would be located at around $160\Mm$ depth ($r=0.78\,R$) according to the mixing length model of
\cite{Spruit:1974} or $r=0.72\, R$ for the model of \cite{Stix:02}.
Therefore, this kind of dynamo wave cannot be driven by the
near-surface shear layer, it must instead be driven by a positive
radial shear and an inversion of sign of $\alphapp$ in the deeper part
of the convection zone to get an equatorward migrating magnetic field
\citep{DWBF15}.

\section{Conclusions}

We use 3D MHD global dynamo simulations to investigate the rotational
dependency of magnetic activity cycles. For moderately and rapidly
rotating runs ($\Co\ge 6.5$), we find well-defined cycles in range between 2 to 4
yrs. For slowly rotating runs, we find irregular cycles with mostly
longer periods. There the cycle periods can be only ill-determined.
Using the $\phi\phi$ component of the $\alpha$ tensor measured with
the test-field method, and the radial shear we find a good agreement
of the cycle period predicted by a Parker-Yoshimura dynamo wave for
moderately to rapid rotating runs.
There we find that the cycle period only weakly depends on rotation
($P_{\rm cycl} \propto \Co^{0.25\pm 0.04}$ and $P_{\rm cycl} \propto P_{\rm
  rot}^{-0.33\pm 0.05}$). Also this scaling is well reproduced by a
Parker-Yoshimura dynamo wave. $\alphapp$ increases only weakly with
rotation ($\alphapp \propto \Co^{0.70\pm 0.25}$) and the strength of negative
radial shear decreases larger than linear with rotation
$\partial\Omega/\partial  r \propto \Co^{-1.33\pm 0.18}$.
This is not in agreement with mean-field theory, where $\alpha$
depends linear on the rotation \citep[e.g.][]{KR80}. Also from models
of differential rotation ones finds the radial shear has not a strong
dependency on rotation \citep{KR99}. However, these models usually
look at the global quantities and we determine our scaling from
localized region responsible for driving the dynamo wave.

Looking at the ratio of rotation and cycle period over Coriolis
number, we do not find any indication of activity branches with
positive slopes as found from observation of stellar cycles by
\cite{BST1998} and \cite{SB1999}.
The negative slope of our simulations with well determined cycles seems to be more
in agreement with the transitional branch postulated by
\cite{DLLMS17} and confirmed by \cite{BMJRCMPW17} and \cite{OLKPG17}.
Furthermore, our results suggest that the cyclic magnetic fields found
in the work by \cite{SBCBN17} are also caused by a Parker-Yoshimura
dynamo wave, because indeed their simulations produce strong negative
shear in the location, where the magnetic field is oscillating.
By assuming the solar magnetic cycle is caused by a Parker-Yoshimura
dynamo wave following a similar scaling as in our simulations, we can
conclude that the dynamo in the Sun operates near the bottom of
convection zone, where turbulent velocities are around $20\mpsec$.

The cycle period dependence on rotation of our simulations can be also well
explained via the rotational quenching of the turbulent diffusivity,
which contribution has to balance with the dynamo driver in the
saturated stage. We find the nearly the same scaling for the turbulent
eddy diffusivity with Coriolis number as expected from a direct inverse
proportionality of the diffusivity with cycle period \citep[e.g.][]{RS72}.

For the slowly rotating runs a Parker-Yoshimura dynamo wave seems to
be not excited, as the predicted periods do not fit with the measured
ones. This might be due to a higher turbulent diffusion in this
rotation regime as found in
\cite{KKB09}. In particular, these simulations need to be further
investigated using the full set of turbulent transport coefficients.
Moreover, we note here that in the rapidly rotating regime the
magnetic field can become highly non-axisymmetric as found recently in
observations \citep[e.g.][]{LJHKH16} and simulations \citep{VWKKOCLB17}.
The work of \cite{VWKKOCLB17} indicate a weaker scaling of $P_{\rm rot}/P_{\rm
  cycl}$ with Coriolis number than what we find in our work and this
might be due to the non-axisymmetric magnetic field solution, which
are suppressed in our work because of the wedge assumption.

We stress here that the cycles determined from our simulation are
directly linked to the magnetic field. Observational stellar cycles
are mostly measured from chromospheric activity variations (Ca II H\&K)
or photometry. These are only proxies of the magnetic field strength and might
not capture all the features of cyclic variations.
Therefore, it would be useful to determine also the integrated
variability caused by the simulated magnetic cycles. Furthermore, we
will in future also investigate how coronal heating and therefore
the X-ray luminosity will depend on the cyclic magnetic field. For
this it is crucial to combine convective dynamo models with a coronal
envelope as done in \cite{WBM11, WKMB12, WKMB13,WKKB16}. This is in
particular important to study the role of helicity connecting the
dynamo active stellar convection zones with stellar coronae, as the
magnetic helicity might play an important role in the heating of
coronae \citep{WCBP17}.

\begin{acknowledgements}
We thank the referee G\"unther R\"udiger and our colleagues Maarit J. K\"apyl\"a, Mariangela Viviani and Jyri J. Lehtinen
for comments on the manuscript and discussion leading to this work.
The simulations have been carried out on supercomputers at
GWDG, on the Max Planck supercomputer at RZG in Garching, in the facilities hosted by the CSC---IT
Center for Science in Espoo, Finland, which are financed by the
Finnish ministry of education. 
.W.\ acknowledges funding by the Max-Planck/Princeton Center for
Plasma Physics and 
 from the People Programme (Marie Curie
Actions) of the European Union's Seventh Framework Programme
(FP7/2007-2013) under REA grant agreement No.\ 623609.
\end{acknowledgements}

\bibliographystyle{aa}
\bibliography{paper}

\end{document}